\documentclass[fleqn,10pt]{SelfArx} 

\definecolor{color1}{RGB}{0,0,90} 
\definecolor{color2}{RGB}{0,20,20} 

\newlength{\tocsep} 
\usepackage{lipsum} 

\usepackage{url}
\usepackage{gb4e}

\JournalInfo{May 12, 2014} 
\Archive{} 

\PaperTitle{Distributed Consensus in Content Centric Networking} 

\Authors{Marc Mosko\textsuperscript{1}*} 
\affiliation{\textsuperscript{1}\textit{Palo Alto Research Center}} 
\affiliation{*\textbf{Corresponding author}: marc.mosko@parc.com} 

\newcommand{\keywordname}{Keywords} 

\Abstract{
We describe a method to achieve distributed consensus in a 
Content Centric Network using the PAXOS algorithm.  Consensus
is necessary, for example, if multiple writers wish to agree on the
current version number of a CCNx name or if multiple distributed
systems wish to elect a leader for fast transaction processing.
We describe two forms of protocols, one using standard CCNx
Interest request and Content Object response, and the second
using a CCNx Push request and response.  We further divide
the protocols in to those using the CCNx 0.x protocol where
Content Object name may continue Interest names and the CCNx 1.0
protocol where Content Object names exactly match Interest names.
}

\begin{document}

\flushbottom 

\maketitle 

\tableofcontents 

\thispagestyle{empty} 


\section*{Introduction} 
\addcontentsline{toc}{section}{\hspace*{-\tocsep}Introduction} 

Distributed consensus is vital for today's networks to provide fast, reliable, and lively services.  The PAXOS algorithm~\cite{lamport1998part,lamport2002paxos}
The present work proceeds along the lines of Multi-Paxos \cite{multipaxos}.  Distributed consensus is expensive, so it is common
to use the distributed consensus to elect a leader who can quickly process transactions.  
Like the Chubby~\cite{Burrows:2006:CLS:1298455.1298487}
distributed lock system, a single system with a renewing lease handles the fast transactions.


\section{PAXOS Overview}
Following Lamport~\cite{lamport2002paxos}, we describe the Multi-Paxos protocol, which is built on top of the Basic Paxos protocol.

In Basic Paxos, a Proposer issues requests to a set of Acceptors in two rounds.  In the first round, the Proposer sends a \textit{prepare} request
with a counter $N$, often taken as a natural number though any type with a total order works.  It sends the \textit{prepare} to a least a majority of
acceptors.  When an acceptor receives a \textit{prepare(N)}, with will respond with an acknowledgement that N is the current maximum.  If the
acceptor had accepted any value $(N_v, V)$ in the past, it will include that information to the Proposer.  The Proposer, if it receives acknowledgements
from a majority of the Acceptors, will proceed to the second round and send an Accept request for $(N, V)$ to the Acceptors.  If there were no previous
$(N_v, V)$, then the Proposer can select any $V$ of its choice.  If the accept the
request, they will send an accept response.  If the Proposer receives a majority of Accept responses, then it knows $(N,V)$ was accepted as
the consensus value.  When an Acceptor accepts a value, it tells a Learner about it.  The Learner will tell other interested systems about the
consensus value.

Multi-Paxos uses a series of iterations of Basic Paxos such that a consensus value $V$ can evolve over time, as $\{V_0, \dots, V_i\}$.
Using Basic Paxos, one can select a single master Proposer.  After it has succeeded in Phase 1, it can submit as many values as it wants 
in Phase 2, so it can begin submitting pairs $\{i, V_i\}$.

The distinction between Proposer, Acceptor, and Learner, following~\cite{lamport2002paxos}, is not exclusive.  In fact, we could call them
just Servers and each is a potential Proposer, and Acceptor, and a Learner.  They contend for the Proposer role and all act as Acceptors
and Learners.

In multi-server settings, one ofter constructs the number $N$ as the tuple $(n, id)$, where $n$ is a natural number and $id$ is
the unique identity of a server.  The $id$ could be administratively assigned, or it could be a cryptographic principal identity such
as the hash of a public key.  Using this system ensures that no two servers will ever issue the same number and that ties are
broken, in this case based on the sort order of the server identities.  One could also use the tuple $(n, priority, id)$, where $priority$
is a server-selected value of its willingness to become the master and could change iteration to iteration.

\section{CCNx Distributed Consensus}

Because cached responses are of little use in an on-line consensus protocol, each content producer
should set the MaxAge of a ContentObject [cite the cache control document] to a small value or even zero.  
A small value would allow some retransmissions within those few milli-seconds.  A zero value would prevent
all cache responses.

We will identify a specific program protecting a specific variable using a consensus group
as the tuple $\{grp, prg, var\}$, where $grp$ is a group name, $prg$ is a program name,
and $var$ is a protected variable (which could be considered an advisory lock for a set of
variables).  In some cases, we need to identify a specific version of a group such that we
know the number of acceptors that constitute a majority.  We use the notation $grpver$
for that version of a $grp$.

\subsection{Proposers, Acceptors, and Learners}
The current Proposer, or master, of a consensus group is elected using distributed consensus where
each contending proposer bids to have its value accepted.  The accepted value determines the master
proposer.  The actual Value  is the name of a CCNx content object that describes the proposer.

The set of Acceptors is maintained as a consensus value.  For a new system to enter as an Acceptor
or be removed by the Proposer if it is non-responsive, is done as a protected variable.  This allows
the Proposer to know what constitutes a majority.

The identity of the Listener(s) (the function could be spread over multiple systems responsible for
different notifications) is maintained as a protected value.  The Acceptors know the identity of the
current Listener and will inform it of accept choices.  The Listener will use the identities of
the acceptor group associated with the given value and notify all Acceptors and Proposers.

\subsection{Individual request/response model}
\begin{figure}
\centering
\includegraphics[width=.8\linewidth, clip, page=1, trim=1.4in 4.5in 5.5in 1.2in]{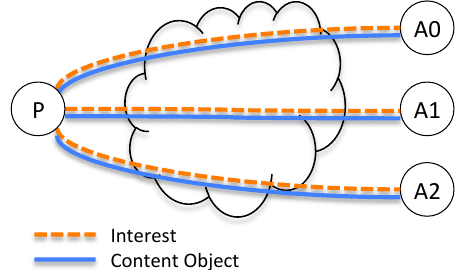}
\caption{Individual Requests}
\label{fig:individual}
\end{figure}

In the individual request/response model, the Proposer must know the identity of
a majority of acceptors.  It can determine this by reading the current value of
the Acceptor group.  The Proposer will send a unicast Interest message to a
majority with a Prepare or Accept request.  The Acceptor will response with a
Content Object response either acknowledging the request or denying the request.
Fig.~\ref{fig:individual} depicts this model.  A proposer $P$ sends individual Interest
messages with payload to a majority of the acceptors, $A_0, A_1, A_2$.
Each accept responds with a Content Object that follows the Interest reverse path.

Any system may read the current consensus value using Eq.~\ref{eq:name0}.  By asking
the Proposer for a \url{read} request, it will return the current consensus value, being the
tuple $(N, tier, V_{tier})$.  A client may also specify a specific $N$ or $(N, iter)$ pair.
If the proposer does not know the consensus value for a read request, it responds with
a NACK.  Following~\cite{lamport2002paxos}, a proposer may also fill in unknown values with
no-ops once it achieves consensus on that value.  A NACK means indeterminate state, while
a no-op means the consensus is there is no value for that state.

\begin{flalign}
& \textrm{\small\url{/proposer/grp/prg/var/read[/N[/iter]]}}\label{eq:name0} & \\
& \textrm{\small\url{/acceptor/grp/prg/var/prepare/N[/iter]}}\label{eq:name1} & \\
& \textrm{\small\url{/acceptor/grp/prg/var/accept/N[/iter]}}\label{eq:name2} & \\
& \textrm{\small\url{/target/grp/prg/var/learn/N[/iter]}}\label{eq:name3} & 
\end{flalign}

A Prepare request message uses the name format shown in Eq.~\ref{eq:name1}.  The \url{/acceptor} prefix
identifies the specific acceptor for routing purposes.  The \url{/grp/prg/var} substring identifies
the consensus group the acceptor participates in, the logical program and protected variable.
The substring \url{prepare} identifies it as a Prepare request.  The suffix \url{/N[/iter]} identifies the
ordering $N$ and the optional iteration $iter$.
If using CCNx 1.0 labeled names, the suffix could take the form of \url{App:prepare = N},
and \url{App:iter = iter}.
The accept request message is similar to the prepare, except for the different identifier
of \url{accept}.

The payload of the request carries the state of the request.  In particular it carries
the value $V$.  Often, the value will be a CCNx 1.0 Link to a particular piece of content
that describes the current state of an algorithm.  In some cases, it links to a CCnx 1.0 Manifest
or is, in fact, that manifest imbedded in the request.

The response of an Acceptor is a CCNx Content Object.  The Content Object follows the
reverse path of the request back to the Proposer.  It carries the consensus state for
the current round or iteration.  Generally, the Content Object should have a short or zero
MaxAge to prevent excessive caching.  Requests for state should go to the Proposer or
the Acceptors to determine what is accepted state.  Fetching a single Content Object response
cannot determine the consensus.

Finally, Eq.~\ref{eq:name3} is used by a Learner.  Acceptors will send an Interest to the
\url{target} of the Learner with an accepted value.  Once the Learner receives a majority
of requests for that state, it can notify other Acceptors and Proposers of the value by
sending the state to them using the same name but with a different \url{target}.  The payload
of the message is the consensus value tuple $(N, [iter,] V_{iter})$.  The Acceptors and Learners
may use semi-reliable communications with Interest out and Content Object ACK back.
If available, the Learner
could use Interest Multicast to Push the learned value, as described in the next section.

Acceptors and Learners may aggregate state in a Learn message.  $N$ would be the
value of the maximum state enclosed in the message, which would then iterate the
learning tuples.

\subsection{Interest multicast model}
\begin{figure}
\centering
\includegraphics[width=.8\linewidth, clip, page=2, trim=1.4in 4.5in 5.5in 1.2in]{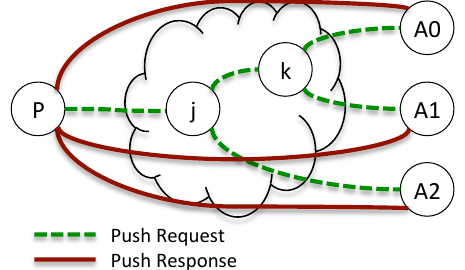}
\caption{Multicast}
\label{fig:multicast}
\end{figure}

Using Interest multicast, a Proposer can send a single Push message to an Interest multicast group and have
all listening acceptors receive the single message.  Because the Proposer needs to know when it has received
at least majority of responses, the group of Acceptors listening to the group name must be identified by a specific
group version with a known number of Acceptors.  Each acceptor needs to send an individual Push response
message back to the Proposer, who identified itself in the Push payload.
Fig.~\ref{fig:multicast} illustrates this mode.  A proposer sends a single Push request to a group name
that uses multicast delivery.  Each acceptor, $A_0 \dots A_2$, responds with an individual Push
response back to the Proposer.  The responses do not necessarily follow the request reverse path.

The multicast model uses similar signaling to the previous model, except the routable prefix is
now the group name \url{grp} rather than an individual member.  In the Learning stage, the
learner may use a mixture of interest/content object exchanges or of multicast push learns.
If using semi-reliable signaling, a node would response with a Push ACK to a learn message.

\begin{flalign}
& \textrm{\small\url{/grp/grpver/prg/var/prepare/N[/iter]}}\label{eq:name1b} & \\
& \textrm{\small\url{/grp/grpver/prg/var/accept/N[/iter]}}\label{eq:name2b} & \\
& \textrm{\small\url{/grp/grpver/prg/var/learn/N[/iter]}}\label{eq:name3b} & 
\end{flalign}

One difference to the individual model is that in some implementation the payload of a request must carry the target
name to use in the response.  This is not strictly necessary because all systems should be aware of
the current system state and know the identity of the Proposer (always the response target for a \url{prepare}
or \url{accept} request) and Learner (always the ACK destination for a Learn message) from the
consensus state and the group version $grpver$.

\section{Conclusion}
We have presented a method to encode and execute the Basic Paxos and Multi-Paxos algorithms
over CCNx 1.0 signaling and messages.  We presented one variation that uses individual
Interest requests and Content Object responses.  We presented a second variation that
uses Push multicast requests and individual Push responses.


\bibliographystyle{unsrt}
\bibliography{ccnxos}


\end{document}